\documentclass{iau} 
\usepackage{graphicx}
\def\mdot{$\dot{M}$}
\title[Evolution of RSGs] 
{The evolution of red supergiants to supernovae}

\author[Emma R. Beasor \& Ben Davies]   
{Emma R. Beasor $^1$ \& Ben Davies $^1$}

\affiliation{$^1$Astrophysics Research Institute, Liverpool John Moores University, Liverpool, L3 5RF, UK \\ email: {\tt e.beasor@2010.ljmu.ac.uk} \\[\affilskip]}

\pubyear{2017}
\volume{329}  
\jname{The Lives and Death-throes of Massive Stars}
\editors{J.J. Eldridge, ed.}
\begin{document}

\maketitle

\begin{abstract}
With red supergiants (RSGs) predicted to end their lives as Type IIP core collapse supernova (CCSN), their behaviour before explosion needs to be fully understood. Mass loss rates govern RSG evolution towards SN and have strong implications on the appearance of the resulting explosion. To study how the mass-loss rates change with the evolution of the star, we have measured the amount of circumstellar material around 19 RSGs in a coeval cluster. Our study has shown that mass loss rates ramp up throughout the lifetime of an RSG, with more evolved stars having mass loss rates a factor of 40 higher than early stage RSGs. Interestingly, we have also found evidence for an increase in circumstellar extinction throughout the RSG lifetime, meaning the most evolved stars are most severely affected. We find that, were the most evolved RSGs in NGC2100 to go SN, this extra extinction would cause the progenitor's initial mass to be underestimated by up to 9M$_\odot$. 
\end{abstract}

\section{Introduction}
Archival imaging now provides a vital tool in identifying the progenitors to supernovae (SNe). Red supergiants (RSGs) end their lives as Type IIP core collapse SNe, of which there have been 7 progenitors confirmed with pre-explosion imaging, most recently the 12.5 $\pm$ 1.2 M$_\odot$ progenitor to SN 2012aw \cite[(Fraser et al. 2016)]{fraser2016disappearence}. Theory predicts that these RSG progenitors should be exploding between masses of 8 to 25M$_{\odot}$ (Ekstr{\"o}m et al. 2012) but so far it seems that the exploding stars are of a relatively low mass with no progenitors appearing in the upper end of the mass range (17 to 25M$_\odot$; \cite[Smartt et al. 2009, 2015]{smartt2009observations,smart2015}).

Since Smartt et al. (2009), various scenarios have been proposed to solve the RSG problem. From a supernova perspective, it was considered whether these high mass RSGs were ending their lives as other types of core collapse supernovae (CCSNe). Smartt et al. (2009) stated that the fractions for type IIn and IIb still did not make up for the lack of high mass RSG progenitors. However, \cite[Smith et al. (2011)]{smith2011observed} disagreed, and suggested that high mass progenitors could indeed be exploding as other CCSNe. 

It is possible that extreme levels of mass loss cause stars to evolve away from the RSG phase and explode as a different class of star. Currently, stellar evolution models rely on observational or theoretical mass loss rate prescriptions, often based on large studies of field stars (e.g. de Jager et al. 1988) or stars that are known to be heavily dust enshrouded (e.g. Van Loon et al. 2005). A potential weakness of using field stars for these studies is that the parameters of initial mass ($M_{\rm initial}$) and metallicity ($Z$) are left unconstrained, possibly causing the large dispersions in the observed trends, while studies targeting heavily dust enshrouded stars are biased towards high-\mdot\ objects.

Georgy  (2012) and Meynet et al. (2015) discussed the implications of increasing these standard \mdot\-prescriptions by factors of 3, 5 and 10 times. These studies found that increasing \mdot\ caused stars to evolve away from the RSG phase at lower masses than predicted by models with standard \mdot, matching the upper mass limit found from progenitor studies. 

There is also the potential that the heaviest RSGs end their lives with no explosion at all. It has been suggested that RSGs with masses higher than 17M$_{\odot}$ may collapse immediately to black hole with little or no explosion (Kochanek et al., 2015). Large observational searches are currently being conducted to find these disappearing stars, with so far only a yellow supergiant as a potential candidate (Reynolds et al., 2015). 
 
However, there could be a simpler solution to the lack of high mass progenitors in the form of circumstellar dust. It has been long established that RSGs form dust in their winds (e.g. de Wit et al. 2008) and it is possible that if a large enough mass of dust built up around the RSG it would appear less luminous, and hence a lower mass would be inferred (as discussed by Walmswell \& Eldridge, 2012). 

To investigate to what degree dust accumulates around the star, and how it might cause the observer to underestimate its initial mass, we have measured the amount of circumstellar material surrounding 19 coeval RSGs, allowing us to investigate whether this is correlated with evolution. 

\section{Application to NGC 2100}
NGC 2100 is a young massive cluster in the LMC rich in RSGs. We assume the cluster is coeval, as any spread in the age of the stars will be small compared to the age of the cluster. This also means the spread in mass between the stars currently in the RSG phase is small, within a few tenths of a solar mass. Using mass tracks and isochrones we find the initial mass for the stars within NGC 2100 to be $\sim$14-17 M$_\odot$ with an age of 15 Myrs. Any difference in luminosity for the RSGs can be considered an evolutionary effect, as the slightly more massive RSGs will evolve at a slightly faster rate, however all the RSGs will follow the same path across the HR diagram. Luminosity can therefore be used as a proxy for evolution. 

\subsection{Modelling results and discussion}
We ran our fitting procedure for 19 RSGs located in NGC 2100. Figure 1 shows the model fit for the most luminous star in our sample with observed photometry. The plot shows our best fit model spectrum (green line), the models within our error range (blue dotted lines) and various other contributions to the total output flux, including scattered flux, dust emission and attenuated flux. It also shows the photometric data (red crosses) and model photometry (green crosses). 
\begin{figure*}
  \caption{ \textit{Left panel:}  Model plot for the star with the highest \mdot\ value in NGC 2100 including all contributions to spectrum.  \textit{Right panel:} Contour plot showing the degeneracy between $\chi^2$ values and best fitting \mdot\ values in units of 10$^{-6}$ M$_\odot$ yr$^{-1}$. The green lines show the best fit \mdot\ and upper and lower \mdot\ isocontours. It can be seen that while there is some degeneracy between inner dust temperature and optical depth the value of \mdot\ is independent of this. Figures taken from Beasor \& Davies (2016).}
  \centering
  \label{fig:allcont}
     \includegraphics[height=5cm,bb=60 0 850 566,clip]{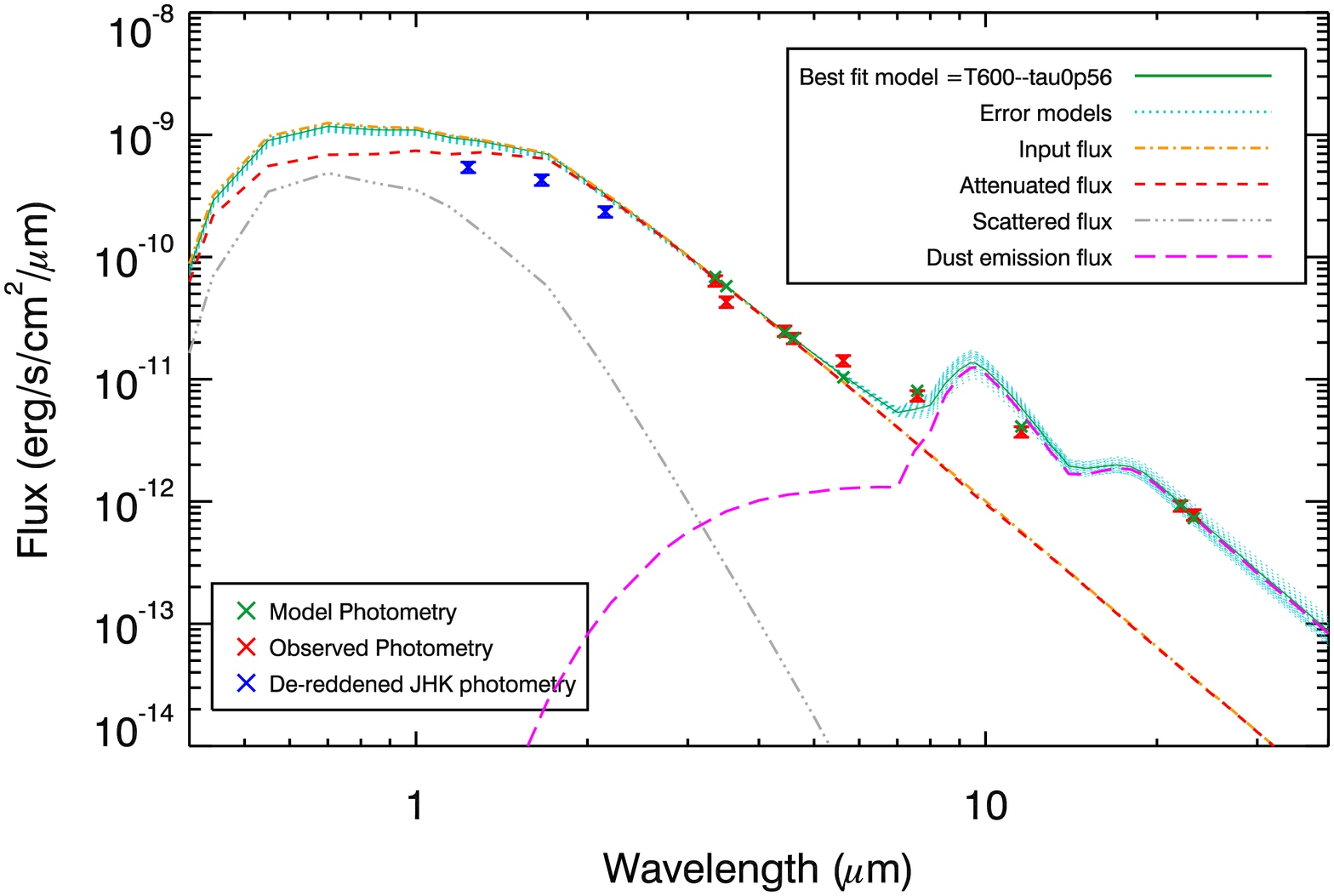}
     \includegraphics[height=5cm]{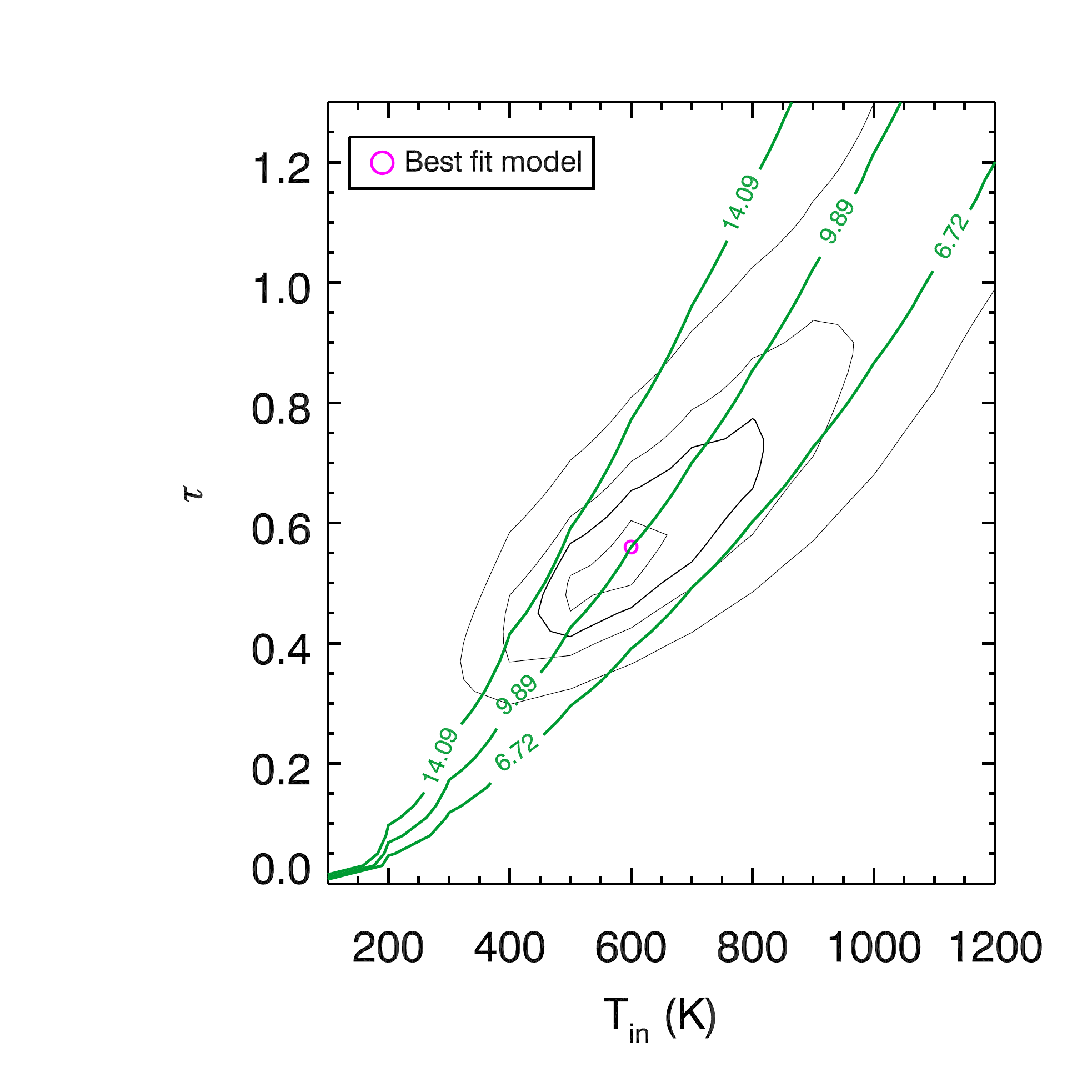}
\end{figure*}

Our model fits allowed us to derive mass loss rates and luminosities for all 19 RSGs in our sample. Figure 2 shows a positive correlation between luminosity and mass loss rate. Since we are using luminosity as a proxy for evolution, the stars with the lowest luminosity can be considered to be the early stage RSGs and the stars with the highest luminosities closest to supernova. Our results suggest \mdot\ increases by a factor of 40 throughout the lifetime of an RSG, approximately 10$^6$ years for a 15M$_\odot$ star (Georgy et al. 2013). Overplotted are commonly used mass loss rate prescriptions. Our correlation is well matched by the prescription of de Jager (de Jager et al. 1988) as it provides the best fit for the more evolved stars (where the mass loss mechanism is stronger). We also find a tight correlation between \mdot\ and luminosity, which we conclude is due to keeping $M_{\rm initial}$ and $Z$ constrained. 

\begin{figure}
  \caption{Plot showing \mdot\ versus $L_{\rm bol}$. A positive correlation can be seen suggesting \mdot\ increases with evolution. This is compared to some mass loss rate prescriptions. The downward arrows show for which stars we only have upper limits on \mdot. Figure taken from Beasor \& Davies (2016).}
  \centering
  \label{fig:8min12}
    \includegraphics[width=80mm]{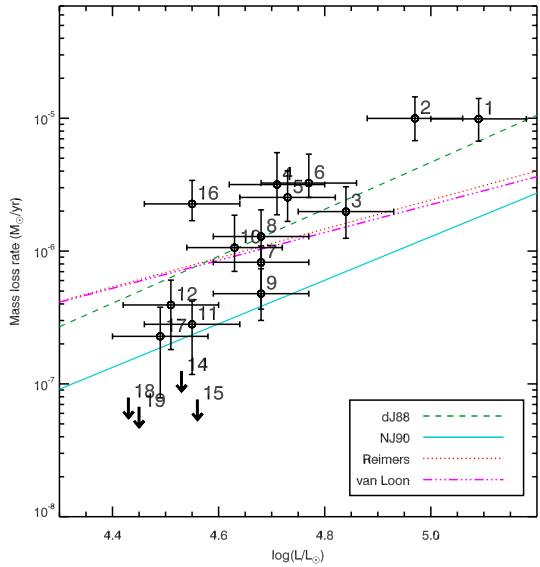}
\end{figure}
We were also able to work out what level of extinction would result from the warm inner dust shell of each star, finding very low levels of extinction that would only have minor effects on mass calculations. However, we did observe an increase in reddening for the two most evolved stars in the cluster. As a check to our model fits, we overplotted JHK photometry (de-reddened for foreground extinction) for all of the stars (which we had not included in our fitting procedure). For the majority of RSGs, the JHK photometry fit the best fit model with no tweaking required. The best fit models for stars \#1 and \#2, the two most evolved stars, instead over predicted the flux at these wavelengths (see Fig. 1 where JHK photometry is shown by the blue crosses). We considered many different scenarios that could be causing this increase in reddening, including assuming an effective temperature that was too high or extreme variability, none of which provided satisfactory solutions (see Beasor \& Davies 2016 for full discussion). 

We next considered the possibility that this increase in reddening was due to cold, clumpy dust at large radii from the stars which would not be detectable with mid-IR photometry. It is known that RSGs have extended, asymmetrical dust shells, a famous example being $\mu$ Cep (de Wit et al. 2008). If we were to move $\mu$ Cep to the distance of the LMC, the cold dust from the extended nebula would be too faint to be observable, at a level of around 0.2 Jy (even before we account for a factor of 2 lower dust to gas ratio in the LMC).  It is therefore plausible that the enhanced extinction we observe for stars \#1 and \#2 is caused by the stars being surrounded by a similar amount of dust that is too faint to detect at the distance of the LMC.
   
\section{Implications}
\subsection{Stellar evolution}
Our results show a clear increase in \mdot\ with RSG evolution, by a factor of $\sim$ 40 throughout the lifetime of the star. For this metallicity and initial mass, we see this \mdot\ is well described by current mass loss rate prescriptions, in particular de Jager's, suggesting there is no need for evolutionary models to increase \mdot\ by significant amounts during the RSG phase. For this $M_{\rm initial}$ ($\sim$14-17 M$_{\odot}$) and at LMC metallicity altering the \mdot\ prescriptions by factors of 10 or more seems unjustified (Georgy 2012, Meynet et al. 2015).

\subsection{Application to SNe progenitors}
We also found evidence for increased reddening to the two most evolved stars in our sample. We now ask the question, if star \#1 were to go SN tomorrow, what would we infer about its initial mass from limited photometric information? Progenitor studies often rely on single-band photometry or upper limits from non-detections, requiring assumptions to be made about spectral type and level of circumstellar extinction. If we apply similar assumptions to those of \cite[Smartt et al. 2009]{smartt2009death} to  \#1, without considering the extra reddening we have observed, we find a mass of 8 M$_{\odot}$. From mass tracks, we have determined the initial mass of the NGC 2100 stars to be 14 - 17 M$_{\odot}$. Hence the mass of the most evolved star in the cluster from single band photometry is clearly underestimated when applying the same assumptions as used by Smartt et al. When we take into account the additional reddening, the mass increases to $\sim$17$\pm$5 M$_{\odot}$, in good agreement with the mass inferred from mass tracks. This is shown in Fig. 3, where the green star represents the mass estimate with no additional extinction considered, the red star is the mass estimate when the additional extinction is considered and the orange line shows the best fit mass track for this cluster. 
\begin{figure}
  \caption{Plot showing the effect of including the additional reddening we observe on progenitor mass estimates for star \#1. The green star represents the mass estimate when no circumstellar extinction is considered, the red star shows the estimated progenitor mass when we take into account additional extinction and the orange line shows the best fit mass track for this cluster. The mass tracks are from Georgy et al. (2013) and are at $Z$ = 0.002.}
  \centering
  \label{fig:LvMDot}
    \includegraphics[height=80mm]{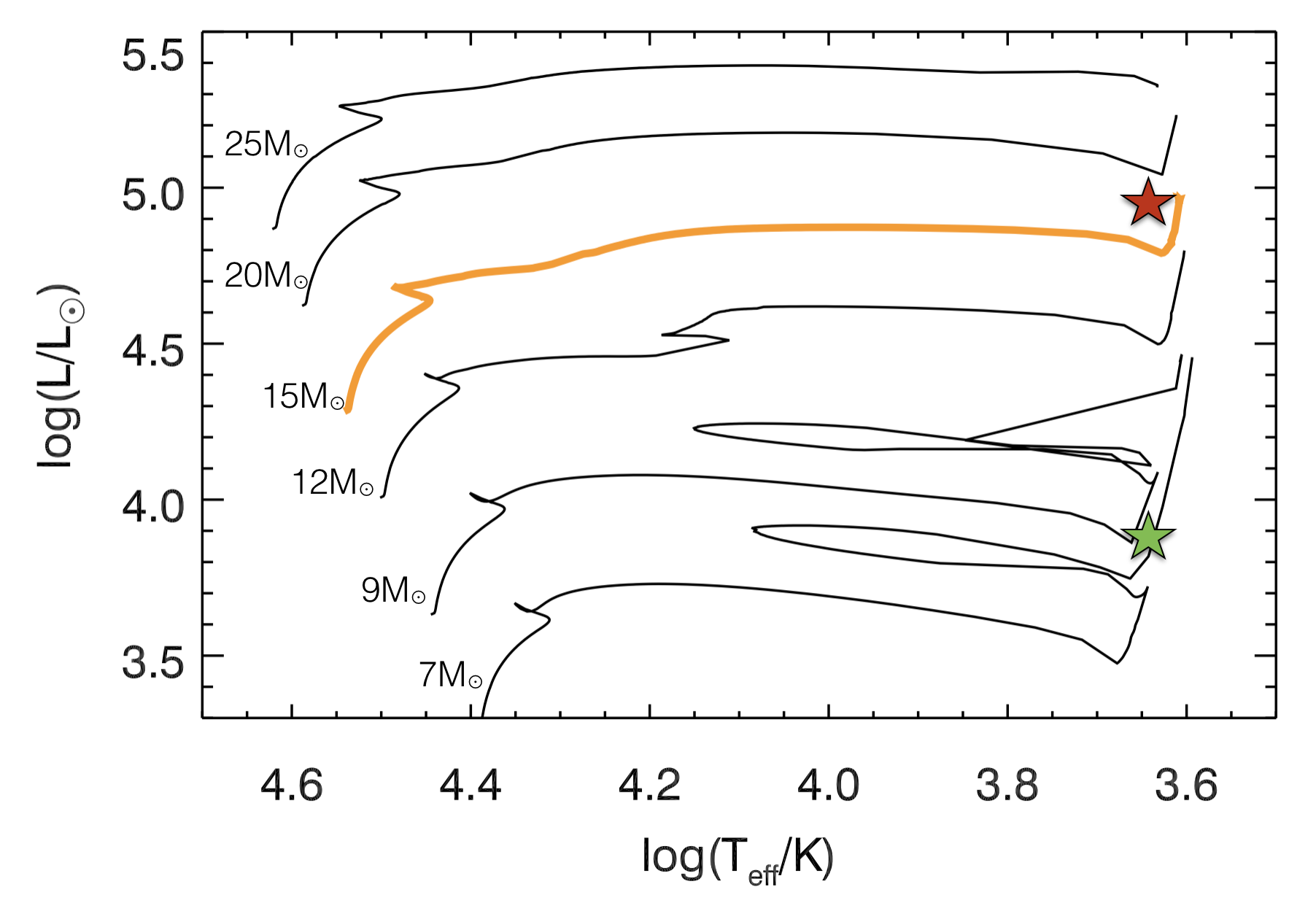}
\end{figure}

With this in mind, we went on to see what effect this level of extinction could have on previously determined progenitor masses. We considered three case studies, the progenitors to SN 1999gi, 2001du and 2012ec (of which SN 1999gi and 2001du are based on upper limits, with SN 2012ec having a detection in one band). The inferred masses of these progenitors increased by 10M$_\odot$, 7M$_\odot$ and $\sim$6M$_\odot$ (see Beasor \& Davies 2016 for full discussion). We have shown that by including similar levels of reddening that we find in the most evolved stars in NGC 2100, the initial mass estimated for Type IIP SNe progenitors increase substantially. If we were to apply this to all objects in the Smartt et al. (2009) sample this may resolve the inconsistency between theory and observations and hence solve the red supergiant problem.


\begin{thebibliography}{}
\bibitem[Beasor \& Davies (2016))]{beasordavies2016}
{Beasor, E., R. \& Davies, B.} 2016, 
\textit{MNRAS}, 463, 1269
\bibitem[]{}
{de Jager, C., Nieuwenhuijzen, H., der Hucht, KA } 1988, 
\textit{Astron. \& Astrophys. Suppl. Series }, 72, 259
\bibitem[]{}
{Fraser, M. et al. } 2016, 
\textit{MNRAS}, 456, L16
\bibitem[]{}
{Ekstr{\"o}m, S., Georgy, C., Meynet, G. et al } 2012, 
\textit{Astron. \& Astrophys.}, 537, A146
\bibitem[]{}
{Georgy, C. } 2012, 
\textit{Astron. and Astrophys.}, 538, L8
\bibitem[]{}
{Georgy, C., Ekstr{\"o}m, S., Eggenberger, P. et al } 2013, 
\textit{Astron. and Astrophys.}, 558, AI03
\bibitem[]{}
{Kochanek, C. S. ,} 2015, 
\textit{MNRAS}, 446, 1213
\bibitem[]{}
{Meynet, G. and Maeder, A.} 2003, 
\textit{Astron. and Astrophys.}, 404, 975
\bibitem[]{}
{Meynet, G. et al} 2015, 
\textit{Astron. and Astrophys.}, 575, A60
\bibitem[]{}
{Reynolds, T., Fraser, M., Gilmore, G} 2015, 
\textit{MNRAS}, 453, 2885
\bibitem[]{}
{Smartt, S., Eldridge, J., Crockett, R., et al.} 2009, 
\textit{MNRAS}, 395, 1409
\bibitem[]{}
{Smartt, S., et al.} 2015, 
\textit{Publ. of the Astron. Soc. of Australia}, 32, e016
\bibitem[]{}
{Smith, N., Li, W., Filippenko, A. et al} 2011, 
\textit{MNRAS}, 412, 1522
\bibitem[]{}
{Walmswell \& Eldridge} 2012, 
\textit{MNRAS}, 419, 2054
\bibitem[]{}
{de Wit, W. et al} 2008, 
\textit{The Astrophys. J. Lett.}, 419, 2054

\end{thebibliography}
\end{document}